\documentclass[12pt]{article}
\usepackage{amsmath}
\usepackage{amsfonts}
\usepackage{amssymb}
\usepackage{graphicx}
\usepackage[round]{natbib}
\usepackage{float}
\usepackage{multirow}
\usepackage{indentfirst}
\usepackage{color, colortbl}
\usepackage{hyperref}
\usepackage{algorithm}
\usepackage{algpseudocode}
\usepackage{setspace} 
\usepackage{multicol}
\usepackage[margin=1in]{geometry}
\newcommand{\vecx}{\mathbf{x}}
\newcommand{\vecz}{\mathbf{z}}
\newcommand{\vecY}{\mathbf{Y}}
\newcommand{\vecy}{\mathbf{y}}
\newcommand{\vecU}{\mathbf{U}}

\newcommand{\vecX}{\mathbf{X}}
\newcommand{\matX}{\mathbf{X}}
\newcommand{\vecA}{\mathbf{A}}
\newcommand{\matZ}{\mathbf{Z}}
\newcommand{\vech}{\mathbf{h}}
\newcommand{\vecvartheta}{\boldsymbol{\vartheta}}
\newcommand{\vectheta}{\boldsymbol{\theta}}
\newcommand{\matSigma}{\boldsymbol{\Sigma}}
\newcommand{\vecmu}{\boldsymbol{\mu}}
\newcommand{\vecdelta}{\boldsymbol{\delta}}
\newcommand{\veczero}{\boldsymbol{0}}
\newcommand{\vecone}{\boldsymbol{1}}
\newcommand{\vecalpha}{\boldsymbol{\alpha}}
\newcommand{\veclambda}{\boldsymbol{\lambda}}

\newcommand{\clustvarsel}{{\tt clustvarsel}}
\newcommand{\vscc}{{\tt vscc}}

\newcommand{\vsccm}{{\tt vsccmanly}}
\newcommand{\skewvarsel}{{\tt skewvarsel}}
\definecolor{Gray}{gray}{0.9}

\usepackage{tabularx}
\usepackage{makecell}
\usepackage{array}
\usepackage{epsfig}
\usepackage{svg}

\usepackage{subfig}

\title{Flexible Variable Selection for Clustering and Classification}
\author{Mackenzie R. Neal and Paul D. McNicholas}
\date{\small Department of Mathematics \& Statistics, McMaster University, Ontario, Canada.}

\begin{document}

\maketitle
\begin{abstract}

The importance of variable selection for clustering has been recognized for some time, and mixture models are well-established as a statistical approach to clustering. Yet, the literature on variable selection in model-based clustering remains largely rooted in the assumption of Gaussian clusters. Unsurprisingly, variable selection algorithms based on this assumption tend to break down in the presence of cluster skewness. A novel variable selection algorithm is presented that utilizes the Manly transformation mixture model to select variables based on their ability to separate clusters, and is effective even when clusters depart from the Gaussian assumption. The proposed approach, which is implemented within the {\sf R} package {\tt vscc}, is compared to existing variable selection methods --- including an existing method that can account for cluster skewness --- using simulated and real datasets.\\[-10pt]

\end{abstract}

\section{Introduction}
Variable selection refers to the process by which informative variables are retained and uninformative variables are removed. Eliminating uninformative variables can improve both model fitting and model interpretability. As such, much research has been conducted on variable selection across statistical domains. One such domain is that of model-based clustering and classification --- in this context, a variable is informative or not depending on whether it is useful for classification or clustering.
The need for dimension reduction is evident for clustering and classification problems as noisy data can hide key features, such as groupings. A popular school of thought is that dimension reduction should happen in tandem with data clustering rather than before clustering \citep{steinley11b,bouveyron14,mcnicholas16b}. As such, variable selection methods that are embedded into clustering and classification algorithms are essential. Many such algorithms exist for Gaussian clustering algorithms; however, the same cannot be said for skewed clustering methods.   
 
In this paper, we study the effect that skewness has on existing variable selection algorithms for classification and clustering and introduce a skewed analogue of the variable selection method VSCC \citep{andrews14}. Implementations of the original VSCC method and the extension introduced herein are available in the \vscc\ package \citep{andrews22} for {\sf R} \citep{R23}.
This extension of VSCC is compared to a skewed extension of the \clustvarsel\ algorithm \citep{raftery06, maugis09a, scrucca18} called \skewvarsel\ \citep{wallace18}, using both real data and simulated data. 

\section{Background} \label{background}
\subsection{Finite Mixture Models}
Finite mixture models arise from the assumption that a population contains sub-populations that can be modelled by a finite number of densities. Thus, these models lend themselves to clustering and classification problems. A random vector $\vecX$ belongs to finite mixture model if the density can be written
$$f(\vecx|\vecvartheta) = \sum_{g=1}^G \pi_gf_g(\vecx|\vectheta_g),$$
where $\pi_g>0$ are the mixing proportions such that $\sum_{g=1}^G\pi_g = 1$ and $f_g(\vecx|\vectheta_g)$ are the component densities. 
Most commonly, these component densities are taken to be multivariate Gaussian corresponding to the following component densities:
$$f(\vecx|\vecmu_g,\matSigma_g) = \sum_{g=1}^G \frac{\pi_g}{(2\pi)^{p/2}|\matSigma_g|^{1/2}}\exp\left\{-\frac{1}{2}(\vecx-\vecmu_g)^{\top}\matSigma_g^{-1}(\vecx-\vecmu_g)\right\},$$
where $\vecmu_g$ denotes the mean and $\matSigma_g$ is the covariance matrix for the $g$th component.
However, in real applications it is uncommon for clusters to be Gaussian; in fact, clusters quite often cannot be well-represented by a Gaussian mixture \citep[see][for discussion]{franczak14,mcnicholas16a}. Thus, various asymmetric mixture models have been developed to aid in clustering and classification when skewness is present (see Section~\ref{smm}). 

\subsection{Skewed Mixture Models} \label{smm}
There are two schools of thought when it comes to dealing with skewness. The first accounts for skewness directly with the use of flexible, asymmetric distributions. These include skew-symmetric distributions such as the skew-normal with density \citep{pyne09}:
$$f(\vecy| \vecmu, \matSigma, \vecdelta) = 2\phi_p(\vecy|\vecmu,\matSigma)\Phi_1(\vecdelta^{\top}\matSigma^{-1}(\vecy-\vecmu);\veczero,\vecone-\vecdelta^{\top}\matSigma^{-1}\vecdelta),$$
where $\phi_p$ and $\Phi_p$ are the pdf and cdf, respectively, of the multivariate Gaussian distribution; $\matSigma$ is the covariance matrix; $\vecdelta$ is the vector of skewness parameters; and $\vecmu$ is the location parameter vector.
Other common asymmetric distributions include the generalized hyperbolic distribution \citep{browne15} and special and/or limiting cases thereof such as the normal inverse Gaussian \citep{karlis09}, variance-gamma \citep{smcnicholas17}, and shifted asymmetric Laplace \citep{franczak14} distributions.  
These distributions are examples of normal variance-mean mixtures. A $p$-dimensional random vector $\vecX$ is a normal variance-mean mixture if its density can be written as 
$$f(\vecx|\vecmu, \matSigma, \vecalpha,\vectheta) = \int_0^\infty \phi_p(\vecx|\vecmu+y\vecalpha,y\matSigma)h(y|\vectheta)dy,$$
where $\phi_p(\vecx|\vecmu+y\vecalpha,y\matSigma)$ is the density of a $p$-dimensional multivariate normal distribution with mean $\vecmu+y\vecalpha$ and covariance $y\matSigma$ and $h(y|\vectheta)$ is a density function for an asymmetric random variable $Y > 0$ \citep{barndorff82}. In Section~ \ref{simresults}, we generate data from a mixture of multivariate variance-gamma distributions to compare variable selection methods in the presence of skewness. Data from a multivariate variance-gamma distribution can be generated via
$$\vecX = \vecmu + Y\vecalpha + \sqrt{Y}\vecU,$$
where $Y\sim \text{gamma}(\lambda,\psi/2)$ and $\vecU\sim N_p(0,\matSigma)$ to result in $\vecX\sim V_p(\lambda,\psi,\vecmu,\matSigma,\vecalpha)$ \citep{smcnicholas17}.

The other school of thought for dealing with skewness utilizes transformations to near-normality. Two transformation mixture models exist: the first is a t-mixture model with a Box-Cox transformation \citep{lo2012}. This model, however, would suffer from the shortcomings of the Box-Cox transformation, primarily its inability to handle left skew \citep{box1964}. Additionally, the Box-Cox t-mixture assumes a global transformation parameter; accordingly, transformations do not vary by variables and components \citep{lo2012}. The second transformation mixture model is a normal mixture model with a Manly transform \citep{zhu18}. The Manly can handle both left and right skewed data, it can be applied to any real number, and it is given by
\[
    T(\vecx|\lambda)= 
\begin{cases}
    \frac{1}{\lambda}\left[\exp\{\lambda \vecx\}-1\right]& \text{if } \lambda\neq0,\\
    \vecx              & \text{otherwise}.
\end{cases}
\]
By applying the back transform of the Manly, one will arrive at the following transformation-based density:
$$f_T(\vecx|\vecvartheta)=\phi(T(\vecx|\veclambda)|\vecmu,\matSigma)J_T(\vecx|\veclambda).$$
where $\vecx$ is the original $p$-dimensional data vector, $\veclambda=(\lambda_1,\lambda_2,\ldots,\lambda_p)^{\top}$ is the transformation vector,
$\vecmu$ is the location parameter vector, and $\matSigma$ is the covariance matrix. The Jacobian of the back-transformation can be written $$J_T(\vecx|\veclambda)=\exp\{\veclambda^{\top}\vecx\},$$ and \citet{zhu18} used this back-transformation to obtain a skewed finite mixture model.
This mixture model contains transformation parameters for each variable-cluster combination. As such, by incorporating the Manly into a model one must introduce ${G}{p}$ additional transformation parameters, potentially resulting in overparameterization. However, \citet{zhu18} recognized that it is unlikely for all variables to need to be transformed in all components. Therefore, to avoid over-parameterization, unnecessary transformation parameters are determined and zeroed out via a backwards or forwards selection process \citep{zhu18}.

Forwards selection begins with a fully Gaussian mixture model (GMM). The GMM is then compared to ${G}  {p}$ models each with one non-zero transformation parameter. The value of this non-zero transformation parameter is selected based on the simplex method, where the conditional expectation of the complete-data log-likelihood is maximized with respect to the skewness parameter in question. For each resulting model, the Bayesian information criterion \citep[BIC;][]{schwarz78} can be obtained via
\begin{equation}\label{eqn:bic}
\text{BIC} = 2\log\hat{L}-p\log n,
\end{equation}
where $\hat{L}$ is the maximized likelihood. Among the $G  p$ candidates, we select the model that maximizes the BIC. The algorithm continues until there are no further improvements to BIC, where parameters from the previous step are used as initializations in the next step.

Backwards selection begins with a fully skewed Manly mixture model. Iteratively, one transformation parameter is zeroed out and the BIC is obtained and compared. Again, this process is continued until no further improvements to BIC are obtained. {Non-zero transition parameters are estimated via the Nelder-Mead algorithm, and remaining model parameters are estimated via an EM algorithm.}
We utilize the aforementioned work of \citet{zhu18} on the Manly mixture to extend the VSCC algorithm into the skewed space, this extension is detailed in Section~\ref{algo}. {However, before doing so we must discuss properties of the Manly component densities that influence clustering results.}

{\subsubsection{Invariance of the Manly Component Densities}
The component Manly densities can be shown to be scale and shifting invariant, thus, operations performed to the data will not have an affect on the clustering results. A common pre-processing operation when clustering is to scale the data, thus such invariance is essential. \citet{zhu18} show that if $\vecX$ is distributed according to $f_\vecX(\vecx; \vecmu, \matSigma, \veclambda) = \phi(\mathcal{M}(\vecx;\veclambda);\vecmu,\matSigma)\exp\{\veclambda^\top\vecx\}$ and $\tilde{\vecX} = \vecA\vecX+\vech$, where $\vecA$ is a diagonal scaling matrix and $\vech$ is a shift vector, then
$$f_{\tilde{\vecX}}(\tilde{\vecx};\vecmu,\matSigma,\veclambda,\vecA,\vech) = f_{\vecX}(\tilde{\vecx};\tilde{\vecmu},\tilde{\matSigma},\tilde{\veclambda}).$$
This then proves invariance, and shows that the inclusion of parameters $\vecA$ and $\vech$ leads to a non-identifiable model. This does not negate the following comments about identifiability.} 

{\subsubsection{Identifiability of the Manly component densities}
A statistical model with parameter space $\Theta$ is identifiable if the following is true,
$$\mathcal{P}_{\theta_1} = \mathcal{P}_{\theta_2} \Rightarrow \theta_1 = \theta_2 \text{ for all } \theta_1\text{,}\theta_2 \in \Theta.$$
Problems with mixture model identifiability regarding the label switching problem are well known and methods exist for handling said problem, such as requiring $\pi_1 > \pi_2 > \cdots > \pi_G $ \citep{redner84, stephens2000}. Thus, the focus in terms of identifiability is focused on the component densities; in the case of the Manly component densities, it is easy to prove identifiability. This is a desirable quality; however, more recently focus has shifted from theoretical identifiability, as defined above, to empirical identifiability. \citet{hennig2023parameters} defines empirical identifiability as the ability to find a consistent sequence of estimators. Due to the use of the Nelder-Mead algorithm \citep{nelder65} for estimation of the transformation parameters, $\veclambda$, we are not able to discuss any consistency results. However, if we assume that the EM and the Nelder-Mead algorithms return maximum likelihood estimates - an assumption that is frequently made - then we can ensure consistency of the resulting estimators, thus ensuring empirical identifiability. That being said, results pertaining to both theoretical and empirical identifiability would be challenged  in the presence of noisy and correlated variables, as discussed in Section~\ref{varsel}.} 

\subsection{Variable Selection} \label{varsel}

{Herein we explore the effect noisy and correlated variables have on identifiability. Increasing correlation between variables has been shown to weaken identifiability \citep{koot2013probabilistic}.  In the case where $\vecX_1 \sim f_X(x; \vecvartheta)$ and $\vecX_2  = \rho\vecX_1+(1-\rho)\matZ$, where $\matZ$ is some random noise and $\text{cov}(\matX_1,\matZ)=0$. The true covariance matrix of $(\matX_1,\matX_2)$ is
$$\matSigma = \begin{bmatrix}
\sigma_1^2 & \rho\sigma_1^2\\
\rho\sigma_1^2 & \rho^2\sigma_1^2 +(1-\rho)^2\sigma_Z^2
\end{bmatrix},$$
leading to non-identifiability. We can further show this with the Manly distribution where $T(\vecX|\veclambda) \sim \mathcal{N}(\vecmu_x,\matSigma_x)$, $T(\matZ|\veclambda) \sim \mathcal{N}(\vecmu_z,\matSigma_z)$, and $\matZ, \vecX$ are independent. Then, the density of $\vecX + \matZ$ is,
$$\frac{1}{(2\pi)^{p/2}|\matSigma_x +\matSigma_z |^{1/2}}\exp\left\{-\frac{1}{2}(\vecx-(\vecmu_x+\vecmu_z))^{\top}(\matSigma_x+\matSigma_z)^{-1}(\vecx-(\vecmu_x+\vecmu_z))\right\}\exp\{\veclambda^{\top}(\vecx+\vecz)\}.$$
Thus removing the identifiability property of the Manly component densities. This provides much motivation for variable selection as many real datasets are plagued with variables following such correlation schemes, see Section~\ref{results}.  Additionally, variables that are completed unrelated to the clustering structure are often captured, these variables would have a large effect on empirical identifiability of any mixture model, often resulting in an over-estimation of the number of components, as seen in Figure~\ref{fig:dimred2}.}

\begin{figure}[H]%
    \subfloat[\centering True clusters from two-component GMM.]{{\includegraphics[width=0.47\linewidth]{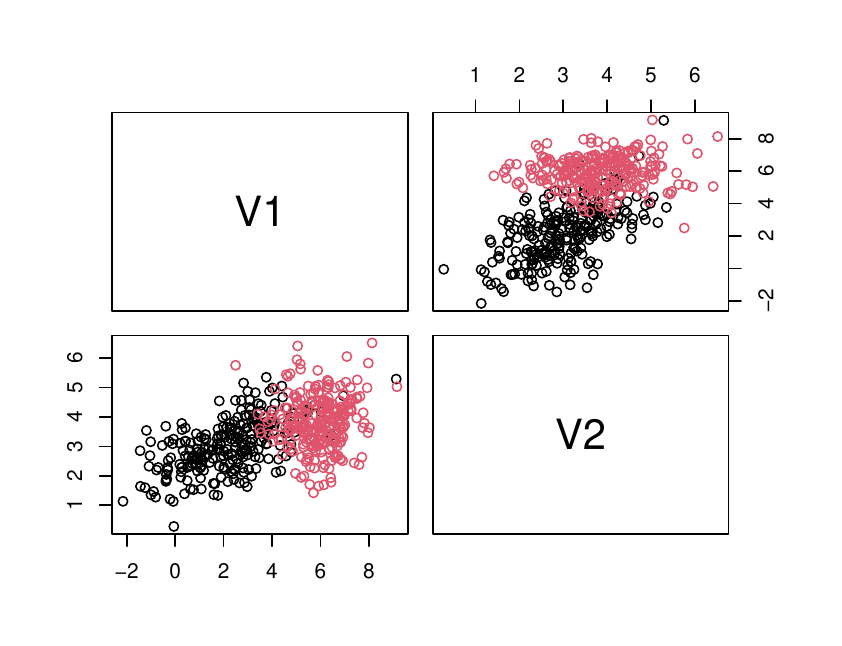}}}%
   \qquad
    \subfloat[\centering Clustering results from GMM fit to simulated data in (a).]{{\includegraphics[width=0.47\linewidth]{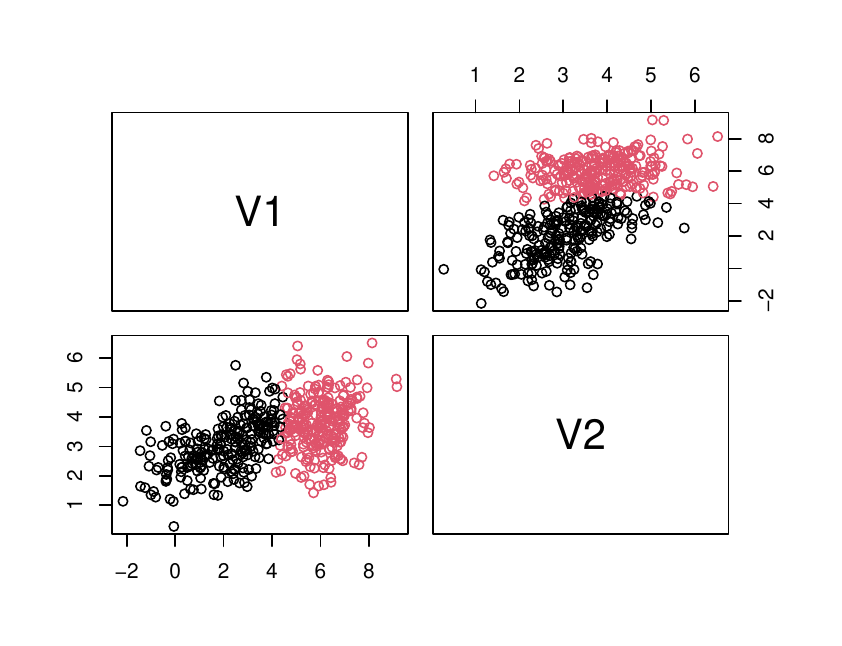}}}%
    \caption{Clustering results from GMM on simulted data.}\label{fig:dimred}
    \end{figure}

\begin{figure}[H]%
\centering\includegraphics[width=0.70\linewidth]{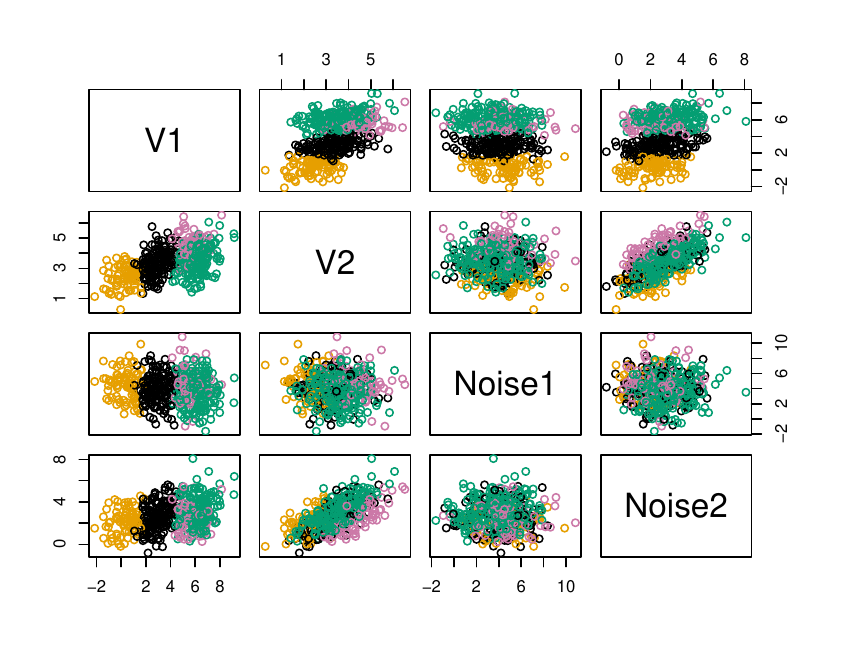}
    \caption{Clustering results from GMM fitted to simulated data in Figure~1(a) plus two noisy variables.}%
   \label{fig:dimred2}%
\end{figure}

As such, the need for dimension reduction algorithms for model-based clustering is clear, this necessity is visualized in Figures~\ref{fig:dimred} and~\ref{fig:dimred2}, where we simulate data from a two-component, two-dimensional GMM and fit a GMM to this data before and after the addition of two noise variables. The first noise variable is random noise generated from a normal distribution, $\text{Noise}_1 \sim N(4,2)$. The second noise variable is correlated to the second clustering variable via $$\text{Noise}_2 = 0.8V2+0.2Z,$$ 
where $Z \sim N(0,5)$ and V2 is a clustering variable. We see that with the addition of just two noise variables, the clustering results begin to break down.

One type of dimension reduction method that could be used to overcome the poor clustering performance seen in Figure~\ref{fig:dimred2} is variable selection. Variable selection is the selection of important variables and the de-selection of unimportant variables --- in the present context, important means important for clustering. Many such variable selection algorithms for clustering and classification exist, and summaries of these algorithms can be found in \citet{steinley08, adams2019, fop18}. We focus on two such algorithms, due to both performance and availability: \clustvarsel\ and \vscc.

\subsubsection{\clustvarsel\ and \skewvarsel}
The \clustvarsel\ algorithm, proposed by \citet{raftery06} and extended by \citet{maugis09a} and \citet{scrucca18}, uses three sets of variables to perform variable selection. The first is the set containing selected variables $\mathbf{X}_{\text{clust}}$, the second is the variable under consideration for inclusion or exclusion $X_{i}$, and the third contains all remaining variables $\mathbf{X}_{\text{other}}$. The Bayes factor \citep{kass95} is used to compare two models essential for variable selection. The first model assumes $X_{i}$ is unimportant for clustering but is related to the set, or a subset, of the clustering variables through linear regression. The integrated likelihood for this model, denoted by $f_1(\mathbf{X}_{\text{clust}},X_i|M_1)$, where $M_1$ is the selected $G$-component GMM, can be decomposed as
$$f_1(\mathbf{X}_{\text{clust}},X_i|M_1) = f_{\text{reg}}(\mathbf{X}_{\text{clust}},X_i)f_{\text{clust}}(\mathbf{X}_{\text{clust}}|M_1),$$
where $f_{\text{reg}}(X_{i}|\mathbf{X}_{\text{clust}})$ is the regression of $X_i$ onto the set, or a subset, of the clustering variables. This subset is selected through stepwise regression, wherein variables from the clustering set are selected if they aid in the prediction of $X_i$. Model one is compared to a second model where $X_i$ is important to clustering and thus the integrated likelihood becomes
$$f_2(\mathbf{X}_{\text{clust}},X_i|M_2) = f_{\text{clust}}(\mathbf{X}_{\text{clust}},X_i|M_2).$$
As discussed by \citet{raftery06} and \citet{maugis09a} the Bayes factor can then be determined as
$$\text{B}_{12} = \frac{f_1(\mathbf{X}_{\text{clust}},X_i|M_1)}{f_2(\mathbf{X}_{\text{clust}},X_i|M_2)}.$$
Because integrated likelihoods are difficult to compute, \cite{kass95} approximate $-2\log\text{B}_{12}$ by
\begin{align*}
 \text{BIC}_{\text{diff}}&=\text{BIC}_{\text{clust}}(\mathbf{X}_{\text{clust}},X_i) - \text{BIC}_{\text{not clust}}(\mathbf{X}_{\text{clust}},X_i)\\
 &= \text{BIC}_{\text{clust}}(\mathbf{X}_{\text{clust}},X_i) - \text{BIC}_{\text{clust}}(\mathbf{X}_{\text{clust}}) - \text{BIC}_{\text{reg}}(X_i|\mathbf{X}_{\text{clust}}).
\end{align*}
Thus, a positive $\text{BIC}_{\text{diff}}$ corresponds to a small Bayes factor, which would suggest that we should cluster on both $X_i$ and $\mathbf{X}_{\text{clust}}$. The \clustvarsel\ algorithm iterates between inclusion and exclusion steps, where one-by-one the variables not in $\mathbf{X}_{\text{clust}}$ are considered for inclusion and variables in $\mathbf{X}_{\text{clust}}$ are considered for exclusion. Variables that maximize $\text{BIC}_{\text{diff}}$ are included and variables that minimize $\text{BIC}_{\text{diff}}$ are removed. As dimensions increase, this algorithm becomes increasingly slow due to its step-wise nature. Additionally, \clustvarsel\ will perform poorly in the presence of skewed clusters due to its reliance on GMMs. {This behaviour is proved rigorously by \citet{loperfido19}, wherein it is shown that the wrongful assumption of symmetric clusters can result in an overestimation of the number of clusters.}

\citet{wallace18} extend \clustvarsel\ into the skewed space with the use of the multivariate skew-normal distribution \citep{pyne09}. The multivariate skew-normal (MSN) is known to be a restrictive asymmetric distribution, being normal-like in the tails, thus making it less robust to outlying observations. Regardless, \citet{wallace18} select the MSN for the skewed extension of \clustvarsel\ due to its computational efficiency, robustness to starting values, and the availability of both regression and mixture model estimation tools, as each are needed in the variable selection implemented in \clustvarsel.

\subsubsection{\vscc}

The \vscc\ algorithm proposed by \citet{andrews14} selects variables based on minimization of within-cluster variance and maximization of between-cluster variance. These goals can be met simultaneously when the data is scaled prior to implementation of the algorithm. The \vscc\ algorithm tends to be much faster than \clustvarsel\ as model fitting is performed on only the original and the final selected variables, rather than at every inclusion/exclusion step. The algorithm begins by calculation of the within-group variance for each variable. The variable that minimizes within-group variance the most is automatically selected into the clustering set. From there, variables are selected into the clustering set based on their ability to separate clusters and their correlation to the set of selected variables. A moving selection criterion is used to do so. This criterion begins with a linear relationship between within-group variance $W_j$ and correlation $\rho_{jr}$ and moves to a quintic relationship. Variable $j$ is selected into the clustering set $V_i$ if, for all $r \in V_i$  the following criteria holds:
$|\rho_{jr}| < 1-W^i_j.$

\begin{figure}[H]
  \centering
  \includegraphics[scale=0.60]{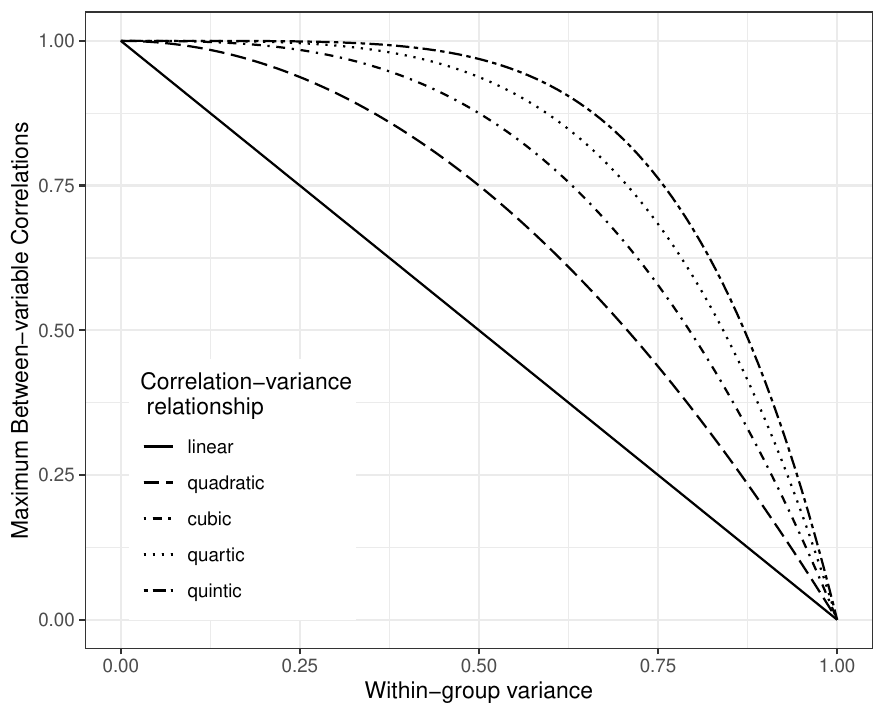}
  \caption{Correlation-variance relationship for \vscc\ selection criteria, based on Figure~1 in \citet{andrews14}.}
  \label{fig:varcor}
\end{figure}

As $i$ increases, the correlation criteria is loosened to allow more correlation between the selected variables. A graphical representation of this relationship, similar to Figure~1 in \citet{andrews14}, is given herein as Figure~\ref{fig:varcor}. The \vscc\ algorithm tests five exponent values $i=1,2,\ldots,5$, resulting in five potential subsets of selected variables. Model-based clustering is carried out on each subset and the final subset is selected based on minimizing the clustering uncertainty \citep{andrews14}, i.e., minimizing $n - \sum_{i=1}^n \max_g\{\hat{z}_{ig}\}$, where 
$\hat{z}_{ig}= 1$ if observation $\vecx_i$ belongs to component $g$, and $\hat{z}_{ig}= 0$ otherwise. Note that minimizing $n - \sum_{i=1}^n \max_g\{\hat{z}_{ig}\}$ is equivalent to maximizing $\sum_{i=1}^n \max_g\{\hat{z}_{ig}\}$.

The \vscc\ algorithm is computationally efficient and performs well on Gaussian clusters. However, because variables are selected based on the minimization of within-cluster variance, this method would suffer substantially when applied to skewed clusters. Herein, we discuss how this algorithm can be extended to skewed data (Section~\ref{method}) and we compare the extension to the previously-discussed algorithms (Sections~\ref{results} and~\ref{simresults}). 

\section{Methodology} \label{method}
\subsection{Algorithm} \label{algo}
We must transform the data to near-normality for minimization of within-cluster variance to be used as a variable selection criterion for skewed clustering/classification problems. Thus, we propose an extension to \vscc\ where a Manly mixture is fit to the data. The transformation parameters are then obtained from the fitted model and applied to the data prior to conducting the variable selection laid out in \vscc. The skewed clustering extension is detailed below in Algorithm~1, where $g=1,\ldots,G$ indexes clusters, $i=1,\ldots,n$ indexes observations, and $j=1,\ldots,p$ indexes variables.

\begin{singlespace}
\begin{algorithm}[H]
\caption{VSCC Manly}
\begin{algorithmic}[1]
\State Perform model-based clustering, fitting either a full Manly mixture or a Manly mixture with transformation parameter selection.
\State Use $\hat{z}_{ig}$ as initial group memberships.
\State Transform data according to the following,
\vspace{-0.075in}
$$\vecY_g=\bigg(\frac{e^{\hat{\lambda}_{1g}\vecx_{1}}-1}{\hat{\lambda}_{1g}},\ldots,\frac{e^{\hat{\lambda}_{pg}\vecx_{p}}-1}{\hat{\lambda}_{pg}}\bigg).$$
\vspace{-0.15in}
\State Scale transformed variables.
\State Calculate within-group variance for each variable, 
\vspace{-0.075in}
$$\hat{W}_j=\frac{1}{n}\sum_{g=1}^G\sum_{i=1}^n\hat{z}_{ig}(y_{ji}-\hat{\mu}_{jg})^2.$$
\vspace{-0.15in}
\State Sort $\hat{W}_j$ in ascending order.
\For{$\texttt{i in } 1:5$}
\State $\hat{W}_1$ is automatically selected into $V_{(i)}$
\For{$\texttt{j in } 2:p$}
\If{$|\rho_{jr}| < 1-\hat{W}^i_j$ for all r in $V_{(i)}$} 
    \State Variable j is placed in $V_{(i)}$
\Else 
  \State Variable j is not placed in $V_{(i)}$ 
\EndIf 
\EndFor
\EndFor
\State Perform Manly mixture model-based clustering on all five variable subsets.
\State Select $V_{(i)}$ such that $n - \sum_{i=1}^n \max_g\{\hat{z}_{ig}\}$ is minimized.
\end{algorithmic}
\end{algorithm}
\end{singlespace}

Algorithm 1 details the skewed extension of \vscc\ for clustering problems. For this method to be applied to classification problems, transformation parameters and true group memberships would need to be supplied in replacement of lines~1 and~2 in Algorithm~1.  

We note that studies of traditional skewed methods versus transformation methods have found that no one type of method for handling skewness outperforms the other \citep{gallaugher20a}; thus, the use of a transformation-based mixture model is an appropriate choice for
dealing with skewness. Additionally, we select a transformation-based mixture model for extending the \vscc\ algorithm into the skewed space, over a mixture model that deals with skewness directly, to allow for transformations of clusters into the near-normal space. 

\subsection{Performance Assessment}
Performance can be easily measured for simulated data as we know the clustering variables \textit{a~priori}. For real data, there are no true clustering variables; as a result, measuring performance becomes more difficult. We measure performance in three ways: adjusted Rand index \citep[ARI;][]{hubert85}, the number of clusters chosen, and visually with variable plots. As the dimension of the selected set increases, it becomes harder to assess performance using visualizations. Regardless, one can still observe redundancy in the selected set and so variable plots remain helpful even in such circumstances. We are operating in the clustering framework; however, true labels exist for all datasets tested. Therefore, the ARI remains a valuable performance assessment tool. 
The ARI was proposed as to force the Rand index \citep{rand71} to have expected value of zero under random assignment \citep{hubert85}, which leads to a more interpretable index. 
The ARI equals one when there is perfect agreement between partitions and is negative when the agreement is worse than would be expected via random assignment.

\subsection{Model Fitting}
All previously discussed methods will be tested on each dataset. To ensure fair comparison between \vsccm\ and \skewvarsel, we fit both a MSN mixture and a Manly mixture to the variables selected by \skewvarsel. An MSN mixture is fitted to remain consistent with \citet{wallace18} and with the \skewvarsel\ algorithm --- the BIC used for variable selection comes from a MSN mixture. 

\section{Real Data Results} \label{results}
The \vscc, \clustvarsel, \vsccm, and \skewvarsel\ algorithms are compared on four datasets under a clustering framework. All methods will test $G=1,\ldots,9$ and data is standardized prior to running each method.

\subsection{Australian Institute of Sport Data}
\begin{figure}[H]
  \centering
  \includegraphics[scale=0.6]{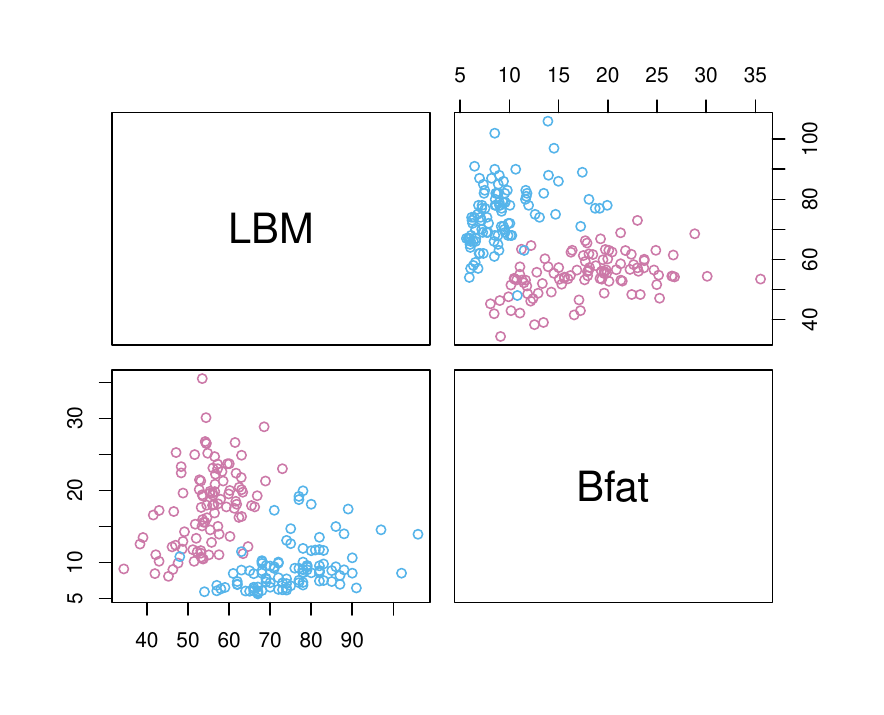}
  \vspace{-0.15in}
  \caption{Variables selected by \vsccm-forwards for the AIS data.}
  \label{fig:vsccmanlyforwardsais}
\end{figure}

\begin{figure}[!h]
  \centering
  \includegraphics[scale=0.7]{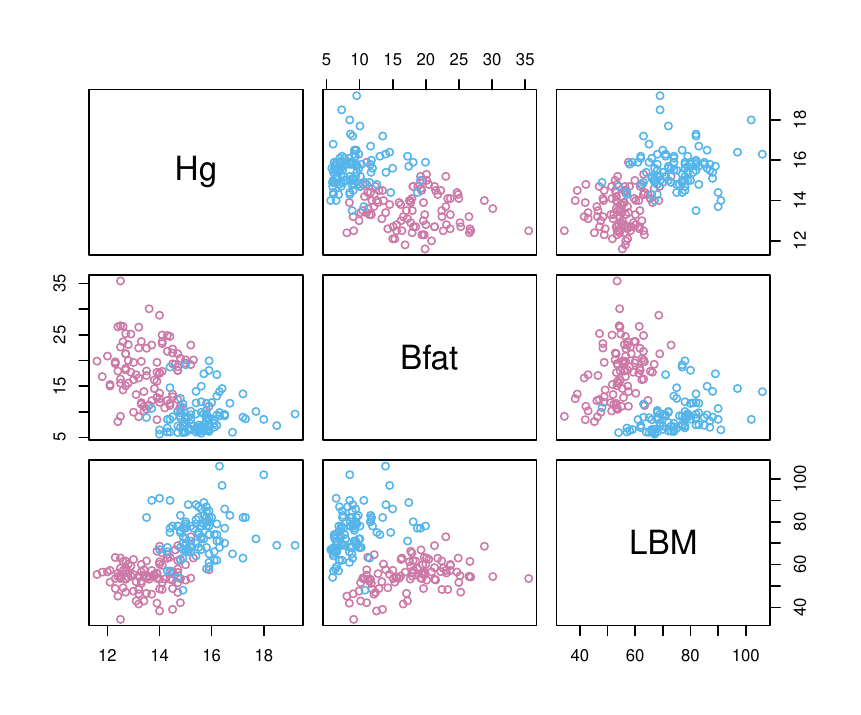}
  \vspace{-0.15in}
  \caption{Variables selected by \vsccm-backwards for the AIS data.}
  \label{fig:vsccmanlybackais}
\end{figure}

The Australian Institute of Sport (AIS) dataset can be found in the {\tt ManlyMix} package \citep{zhu18a}. This dataset contains 11 measurements on 202 individuals. Clustering results are compared to the sex column.  
From Table~\ref{tab:aistab}, we find that the \vsccm\ algorithms perform the best in terms of $G$ and ARI. More significantly, the \vsccm-forwards algorithm reduces the dimensions more than all other methods tested. 
From Figures~\ref{fig:vsccmanlyforwardsais} and~\ref{fig:vsccmanlybackais}, we see that the variables selected by the \vsccm\ algorithms clearly separate the sexes. All other methods tested appear to be more susceptible to correlated variables, thus creating redundancy in the selected set. The \vsccm-forwards and \vsccm-backwards algorithms resulted in the selection of a different final set of variables. Just as forwards and backwards step-wise regression can result in different results, forwards and backwards transformation parameter selection can result in different transformed spaces. Hence, it is unsurprising to see a difference in the set of selected variables.
\begin{singlespace}
\begin{table}[H]
\centering
\caption{Variables selection results for the AIS data.}
    \label{tab:aistab}
\begin{tabular}{l c c p{2in}} 
 \hline
 Model & $G$ & ARI & Variables \\ [0.5ex] 
 \hline
 \vscc & 4 & 0.52 & {LBM, Bfat, SSF, Wt, Ht, Hg, BMI, RCC, Fe }\\ 
 \clustvarsel & 7 & 0.27 & LBM, Bfat, Wt \\
 \rowcolor{Gray}
 \vsccm-forward & 2 & 0.94 & LBM,  Bfat \\
  \rowcolor{Gray}
 \vsccm-backward  & 2 & 0.96 & LBM, Bfat, Hg\\
  \rowcolor{Gray}
 \vsccm-full & 2 & 0.96 & LBM, Bfat, Hg\\
 \skewvarsel\ + MSN & 3 & 0.26 & LBM, Bfat, SSF, Wt\\ 
 \skewvarsel\ + Manly forward & 4 & 0.59 & LBM, Bfat, SSF, Wt \\ 
 \skewvarsel\ + Manly backward & 4 & 0.57 &  LBM, Bfat, SSF, Wt \\ 
 \hline
\end{tabular}
\end{table}
\end{singlespace}
\begin{figure}[!h]
  \centering
  \includegraphics[scale=0.75]{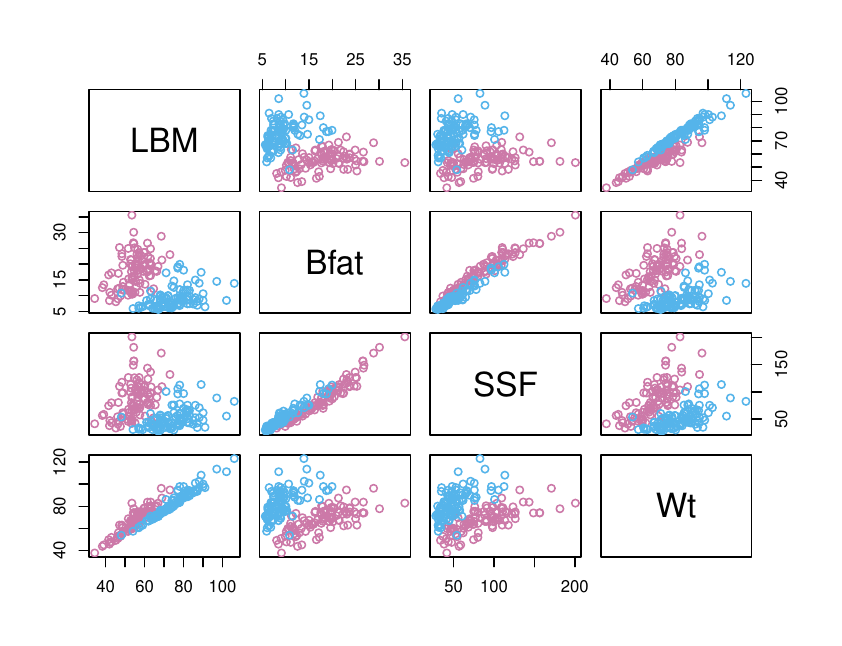}
  \vspace{-0.15in}
  \caption{Variables selected by \skewvarsel for the AIS data.}
  \label{fig:skewvarais}
\end{figure}

\subsection{Banknote Data}
\begin{figure}[H]
    \centering
    \includegraphics[scale=1]{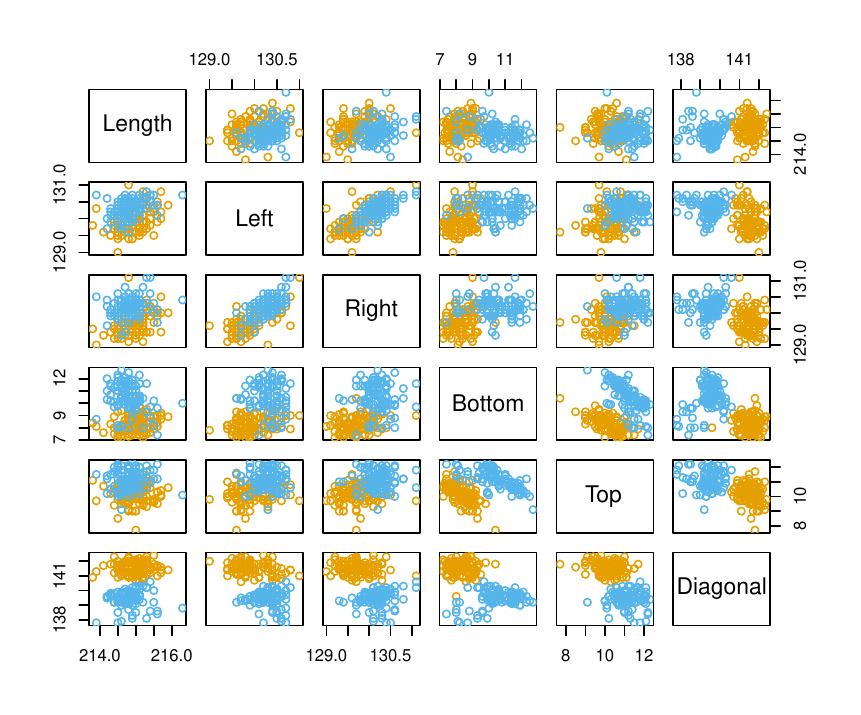}
  \vspace{-0.15in}
    \caption{Variables in the banknote data.}
    \label{fig:bankplot}
\end{figure}

The banknote dataset comes from the {\tt mclust} package \citep{scrucca16}. There are six measurements, 200 observations, and two types of bills (genuine and counterfeit). Clustering results are compared to bill type.  
Variable selection on the banknote dataset produces interesting results as no method significantly reduces dimensions (Table \ref{tab:banktab}). This is surprising because, from Figure~\ref{fig:bankplot}, it appears as though only two variables would be necessary for separating clusters. However, when a Manly mixture is fit to either the variables selected by the \skewvarsel\ or the \vscc\ algorithm, three clusters are found and ARI drops to 0.85.  This suggests that although it may seem like the \vsccm\ algorithm is selecting too many variables, the algorithm may be selecting the number of variables needed to ensure superior clustering performance. 
\begin{singlespace}
\begin{table}[H]
\centering
\caption{Variable selection results for the banknote data.}
\label{tab:banktab}
\begin{tabular}{l c c l} 
 \hline
 Model & $G$ & ARI & Variables \\ [0.5ex] 
 \hline
 \vscc & 3 & 0.86 & Diagonal, Bottom, Top, Right \\ 
 \clustvarsel & 4 & 0.67 & Diagonal, Bottom, Top, Left, Length \\
  \rowcolor{Gray}
 \vsccm-forward & 2 & 0.98 & Diagonal, Bottom, Top, Right, Left \\
  \rowcolor{Gray}
 \vsccm-backward & 2 & 0.98 & Diagonal, Bottom, Top, Right, Left \\
  \rowcolor{Gray}
 \vsccm-full & 2 & 0.98 & Diagonal, Bottom, Top, Right, Left\\
 \skewvarsel\ + MSN & 4 & 0.69 & Diagonal, Bottom, Top, Left\\ 
 \skewvarsel\ + Manly forward & 3 & 0.85 & Diagonal, Bottom, Top, Left \\
 \skewvarsel\ + Manly backward & 3 & 0.85 &  Diagonal, Bottom, Top, Left \\
 \hline
\end{tabular}
\end{table}
\end{singlespace}

\subsection{Italian Wine Data}
The Italian wine dataset can be found in the {\tt pgmm} package \citep{mcnicholas2022}. It contains 28 variables, 178 observations, and three types of wine to which clustering results are compared.  
In Table~\ref{tab:winetab}, we see that \skewvarsel\ performs the best while \vsccm\ performs the worst, in terms of $G$ and ARI. Nearly all methods appear to reduce dimensions approximately the same amount with key variables such as flavanoids and hue being selected nearly every time. This would suggest that it is not the minimization of within-cluster variance that is performing poorly on this dataset but rather the fit of the Manly mixture. This point is further emphasized when we look at the \skewvarsel\ results in Table~\ref{tab:winetab}. When the MSN mixture is fit to the \skewvarsel\ selected variables, a much higher ARI is obtained than when the backwards Manly mixture is fit to the same variables. These results suggest that the Manly distribution may be more prone to combining Gaussian clusters to create skewed clusters.  As the MSN distribution is normal-like in the tails, the MSN may be less prone to the same behaviour. We see this behaviour in the pairs plots of the Italian wine data selected by  \vsccm-backwards (Figure~\ref{fig:wineback}). We replicate this behaviour on simulated data from a three-component, two-dimensional GMM in Figure~\ref{fig:simback}. 
\begin{singlespace}
\begin{table}[H]
\centering
\caption{Variable selection results from the Italian wine dataset.}
\label{tab:winetab}
\begin{tabular}{l c c p{3.2in}} 
 \hline
 Model & $G$ & ARI & Variables \\ [0.5ex] 
 \hline
\vscc & 3 & 0.76 & {Flavanoids, Hue, OD280/OD315 Diluted Wine, Proline, Colour Intensity, Alcohol,Total Phenols, OD280/OD315 of Flavanoids, Uronic Acids, Tartaric Acid, Fixed Acidity, Glycerol, Malic Acid, Alcalinity of Ash, Sugar-free Extract, Proanthocyanins, Phosphate, Non-flavanoid, 2-3-Butanediol, Calcium, Ash, Magnesium, Total Nitrogen, Chloride} \\ 
 \clustvarsel & 5 & 0.66 &  {Flavanoids, Proline, Colour Intensity, Uronic Acid, Chloride, Malic Acid}\\
  \rowcolor{Gray}
 \vsccm-forward  & 2 & 0.43 & {Flavanoids, Hue, OD280/OD315 Diluted Wine, OD280/OD315 Flavanoids} \\
  \rowcolor{Gray}
 \vsccm-backward  & 2 & 0.49  &  {Flavanoids, Hue, OD280/OD315 Diluted Wine, Colour Intensity,  Uronic Acid}\\
  \rowcolor{Gray}
 \vsccm-full &2 & 0.47 & {Flavanoids, Hue, OD280/OD315 Diluted Wine, Colour Intensity, Uronic Acid, Total Phenols}\\
 \skewvarsel\ + MSN & 3 & 0.80 &{Flavanoids, Hue, Proline, Colour Intensity, Alcohol, Uronic Acid, Malic Acid, Tartaric Acid} \\ 
 \skewvarsel\ + Manly forward & 3& 0.73 & {Flavanoids, Hue, Proline, Colour Intensity, Alcohol, Uronic Acid, Malic Acid,Tartaric Acid }\ \\ 
  \skewvarsel\ + Manly backward & 2 & 0.46&  {Flavanoids, Hue,  Proline, Colour Intensity, Alcohol, Uronic Acid, Malic Acid, Tartaric Acid }\ \\ 
 \hline
\end{tabular}
\end{table}
\end{singlespace}

\begin{figure}[H]
  \centering
  \includegraphics[scale=0.80]{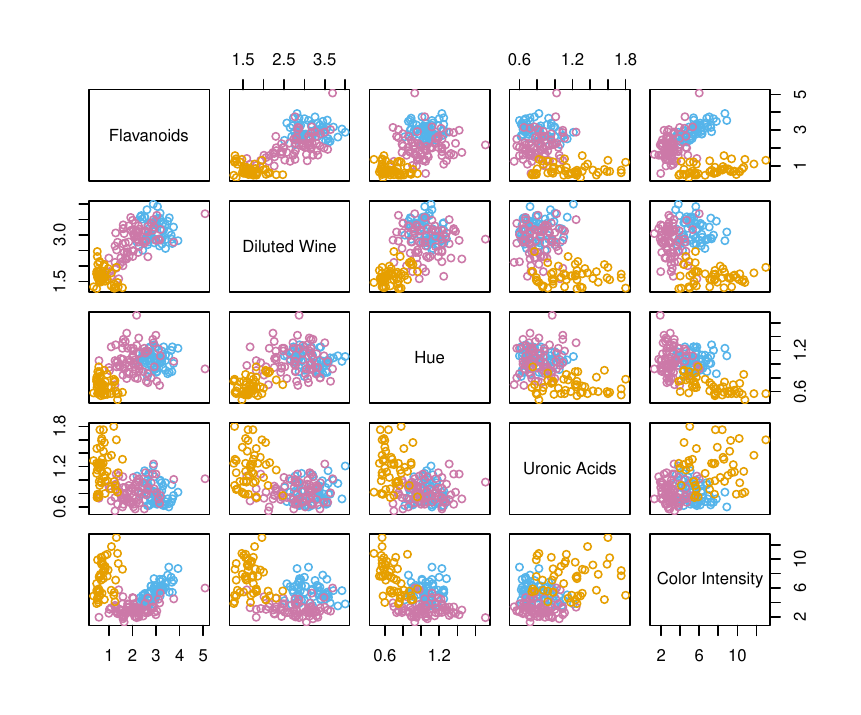}
  \vspace{-0.15in}
  \caption{True clustering on variables selected by \vsccm-backwards for the wine data.}
  \label{fig:truecluswine}
\end{figure}
\begin{figure}[H]
  \centering
  \includegraphics[scale=0.80]{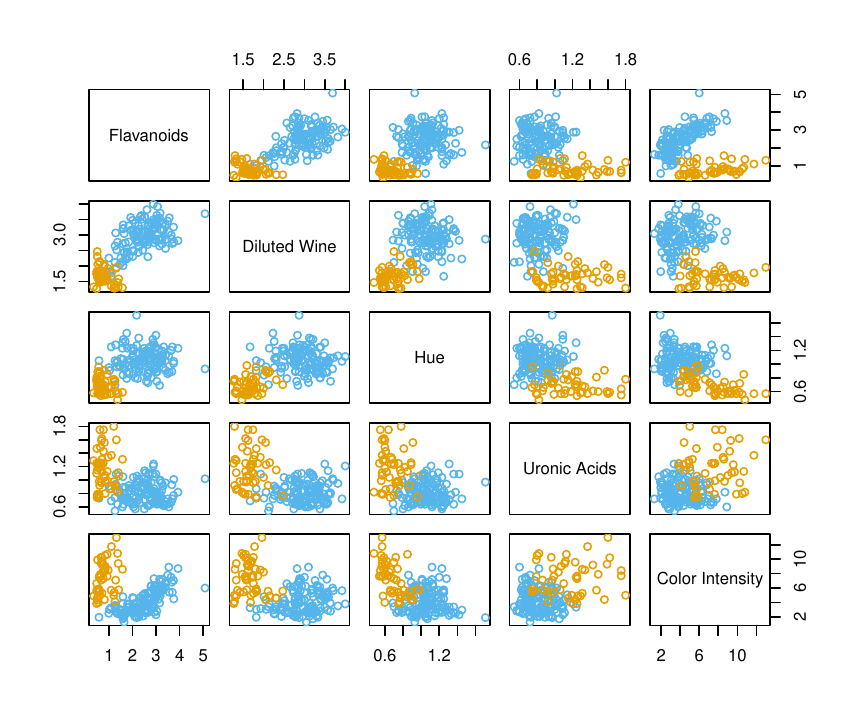}
  \vspace{-0.15in}
  \caption{Clustering by \vsccm-backwards  for the wine data.}
  \label{fig:wineback}
\end{figure}

\begin{figure}[H]%
    \centering
    \subfloat[\centering True clusters (components) simulated data from three-component GMM.]{{\includegraphics[width=0.47\textwidth]{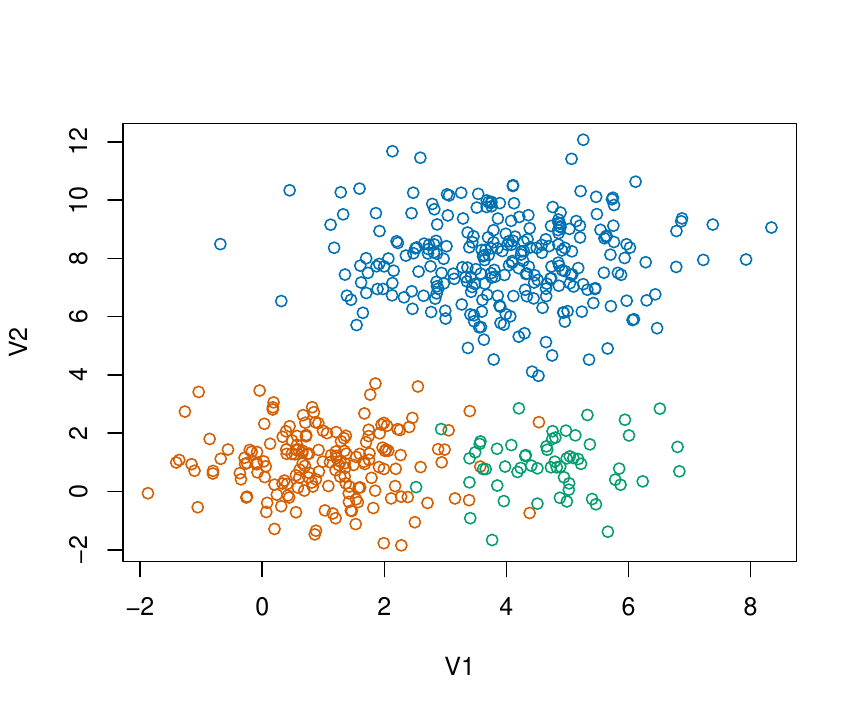}}}%
    \subfloat[\centering Clustering results for \vsccm-forwards.]{{\includegraphics[width=0.47\textwidth]{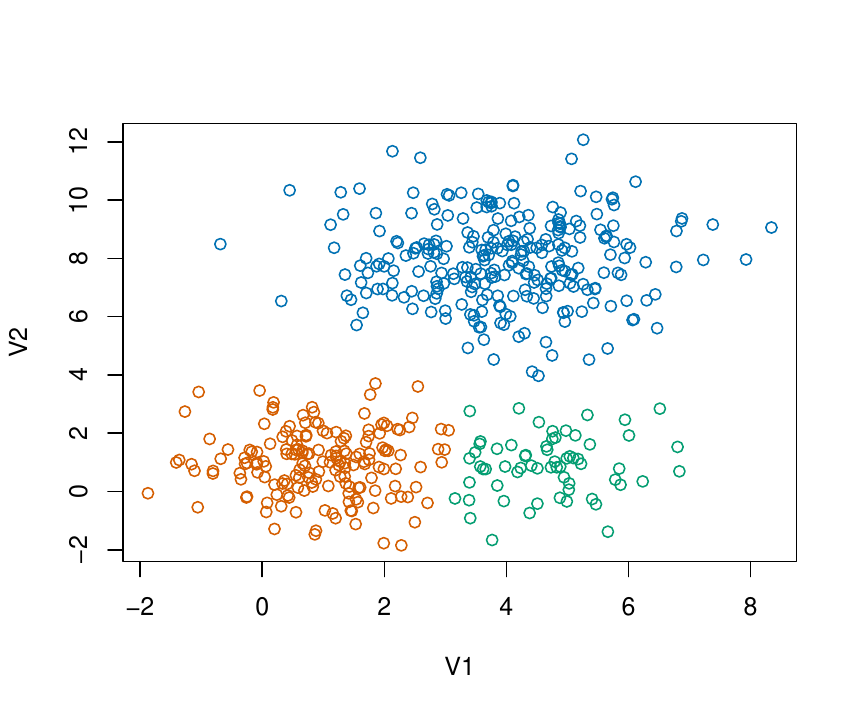}}}%
    \qquad
    \subfloat[\centering Clustering results for \vsccm-backwards]{{\includegraphics[width=0.47\textwidth]{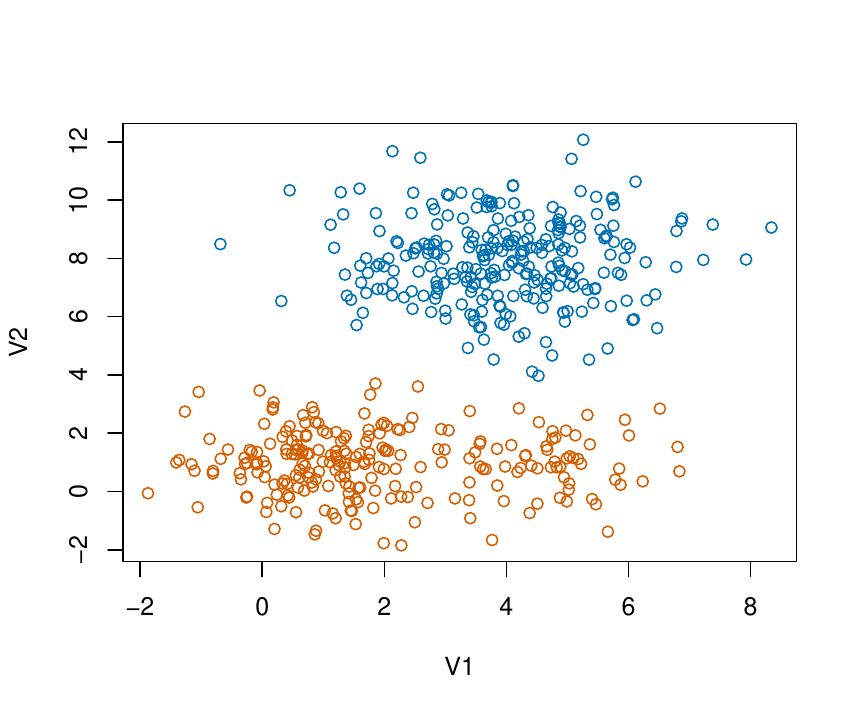}}}%
    \subfloat[\centering Clustering results for \skewvarsel+MSN.]{{\includegraphics[width=0.47\textwidth]{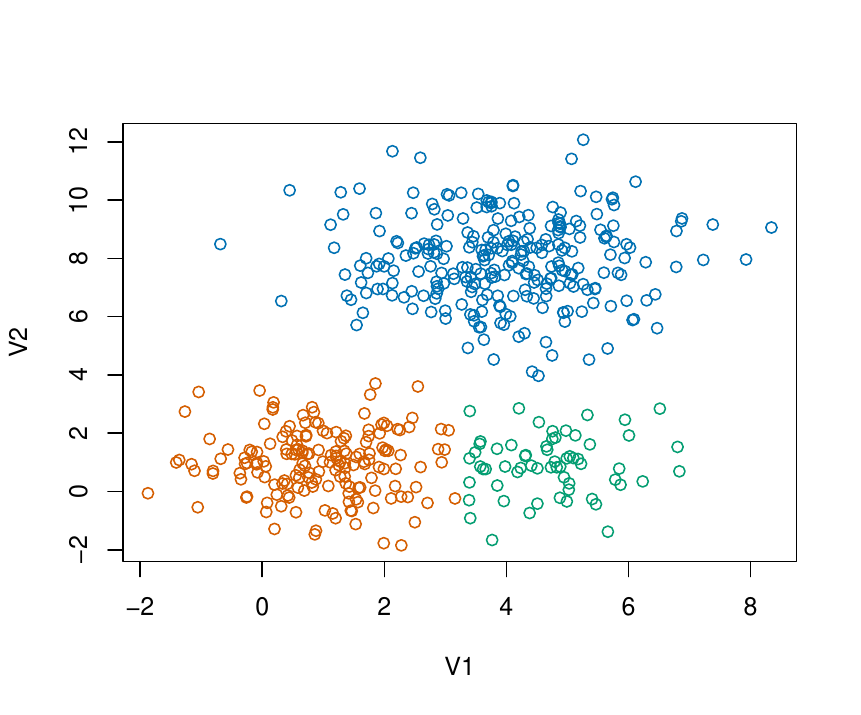} }}%
    \caption{Clustering results from simulated three-component GMM.}%
    \label{fig:simback}%
\end{figure}

\subsection{Breast Cancer Wisconsin (Diagnostic)}
The breast cancer dataset comes from the UCI Machine Learning Repository \citep{Dua2019}. It contains 30 variables, two tumour types (benign and malignant), and 569 observations. Clustering results are compared to tumour type. From Table~\ref{tab:breasttab}, we see that \vsccm\ reduces the dimensions from 30 variables down to one, while selecting the correct number of clusters and obtaining the highest ARI of all methods tested. In particular, we see a large jump in performance, on all three measures, from \vscc\ to its skewed counterpart. We do not see similar improvement in performance by the skewed extension of \clustvarsel. Again, we see redundancy in the variables selected by \skewvarsel, as seen in Figure~\ref{fig:breast1}.
\begin{singlespace}
\begin{table}[H]
\centering
\caption{Variable selection results for the breast cancer data.}
\label{tab:breasttab}
\begin{tabular}{l c c p{3.2in}}
 \hline
 Model & $G$ & ARI & Variables \\ 
 \hline
 \vscc & 9 & 0.16 & V7, V8, V9, V10, V11, V12, V13, V14, V17, V18, V19, V20, V21, V22, V26, V28, V29, V30\\ 
 \clustvarsel & 4 & 0.39 & V3, V5, V6, V8, V9, V13, V15, V16, V18, V19, V22, V23, V25, V26, V28, V29\\ 
  \rowcolor{Gray}
 \vsccm-forward & 2 & 0.30 &  V10, V21\\
  \rowcolor{Gray}
 \vsccm-backward & 2 & 0.63 &  V30\\
  \rowcolor{Gray}
  \vsccm-full & 2  & 0.50 &  V16\\
  \skewvarsel + MSN & 5 &0.33& V3, V6, V13, V16, V23, V26\\ 
 \skewvarsel + Manly forward & 4 & 0.45 & V3, V6, V13, V16, V23, V26\\ 
  \skewvarsel + Manly backward &3  & 0.36 & V3, V6, V13, V16, V23, V26\\ 
 \hline
\end{tabular}
\end{table}
\end{singlespace}

\begin{figure}[H]%
    \subfloat[\centering Variables selected by \vsccm-full.]{{\includegraphics[width=0.47\linewidth]{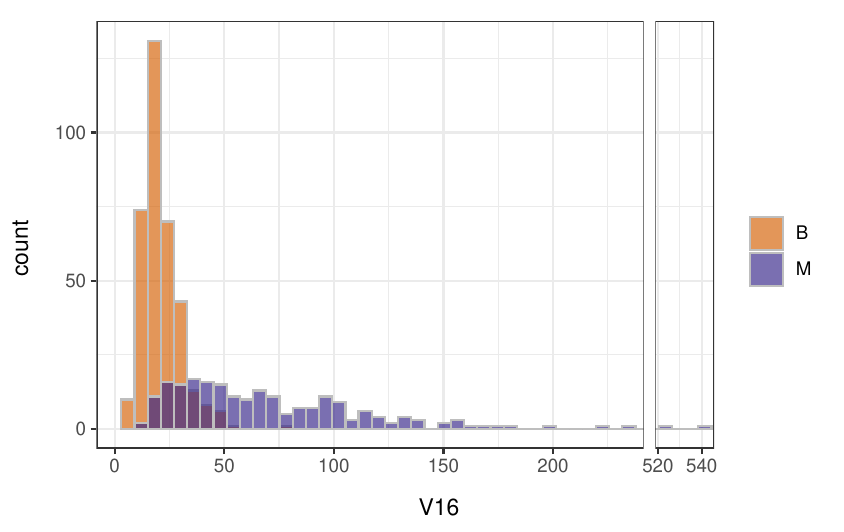}}}%
    \qquad
    \subfloat[\centering Variables selected by \vsccm-backwards.]{{\includegraphics[width=0.47\linewidth]{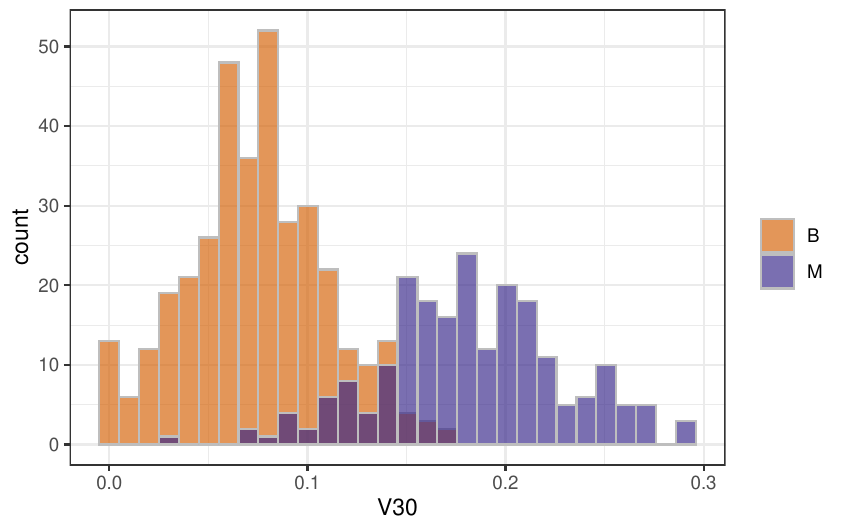}}}%
    
    \caption{Breast cancer variables selected by \vsccm.}%
    \label{fig:breastplots}%
 
\end{figure}

\section{Simulated Data Results} \label{simresults}
We simulated data from a three-component mixture of multivariate variance-gamma distributions 250 times. An example of this data can be found in Figure~\ref{fig:simulatedD}. To allow us to test for the effect of sample size on method performance, we ran our simulation for $\text{N}\in\{200, 500, 1000\}$. Using a simulation is helpful as we can artificially create clustering and non-clustering variables to determine how well these methods select important variables and deselect unimportant ones. The simulation specifics are detailed in Table~\ref{tab:simInfo}, where information on the clustering variables (V1 and V2), nonsense variables (V3 and V4) and the noisy variable (V5) can be found. To reduce the computational time, each model is fit to each simulated dataset only once.
\vspace{-0.2in}
\begin{figure}[H]
    \centering
    \includegraphics[width=0.80\linewidth]{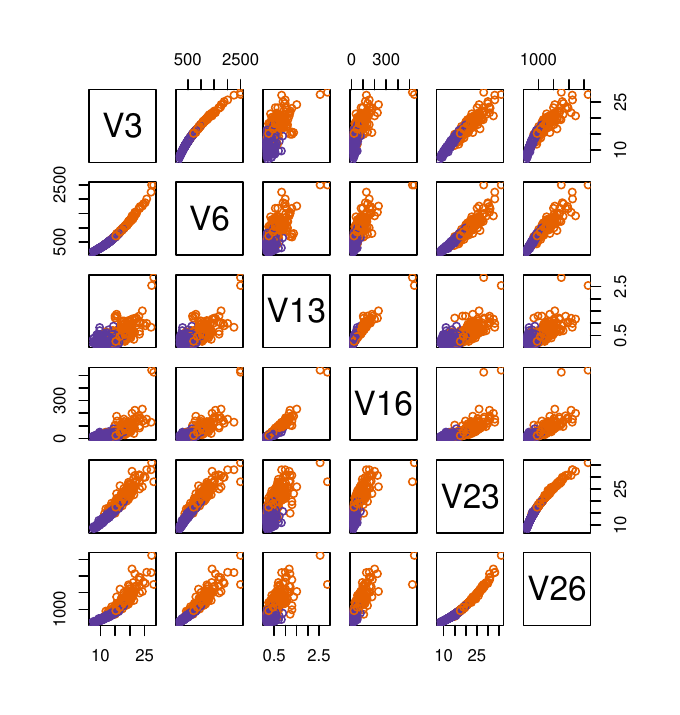}
    \vspace{-0.2in}
    \caption{Breast cancer variables selected by \skewvarsel.}
    \label{fig:breast1}
\end{figure}

From Table \ref{tab:simres}, we see that the \vsccm-backwards, \vsccm-full, and \skewvarsel\ algorithms perform the best in terms of average $G$, average ARI, selecting the correct variables (V1 and V2) and removing the nonsense variables (V3 and V4) when $N=500$ and $N=1000$. However, we see a considerable standard deviation of the ARI when $N=200$ for \skewvarsel, suggesting a possible instability at smaller sample sizes. The \skewvarsel\ algorithm performs the best on removing the noisy variable (V5) across all sample sizes. Generally, these results agree with the real data results in that the skewed methods greatly improve dimension reduction over their Gaussian counterparts when skewness is present. Unlike the real data results, we see the \skewvarsel\ algorithm performing the best, in terms of dimension reduction, on the simulated data suggesting that this algorithm may suffer in the presence of outliers or deviations from theoretical distributions. 
\begin{singlespace}
\begin{table}[H]
\centering
\captionof{table}{Information on how the simulated data are generated.}
\label{tab:simInfo}
\begin{tabular}{ lcccc }
 \hline
Variable Type & \multicolumn{4}{c}{Specifications}\\ \hline

 \multirow{8}{*}{Clustering}&\multicolumn{4}{c}{$[\vecX_{1g},\vecX_{2g}]\sim \text{MVG}(\vecmu_g,\matSigma_g,\vecalpha_g,\lambda_g,\psi_g)$}\\ 
 
&&$\vecmu_1=\begin{bmatrix}
2 \\ 3 
\end{bmatrix}$ & $\vecmu_2=\begin{bmatrix}
5 \\ 3 
\end{bmatrix}$ &$\vecmu_3=\begin{bmatrix}
5 \\ 15 
\end{bmatrix}$\\
 
 &&$\matSigma_1=\begin{bmatrix}
1 & 0 \\
0 & 1 
\end{bmatrix}$ & $\matSigma_2=\begin{bmatrix}
1 & 0 \\
0 & 1 
\end{bmatrix}$ & $\matSigma_3=\begin{bmatrix}
2 & 0 \\
0 & 2 
\end{bmatrix}$ \\
 
&&$\vecalpha_1=\begin{bmatrix}
1 \\ 4 
\end{bmatrix}$ & $\vecalpha_2=\begin{bmatrix}
4 \\ 4 
\end{bmatrix}$ &$\vecalpha_3=\begin{bmatrix}
0.1 \\ 0.1 
\end{bmatrix}$\\
 
 && $\lambda_1=4$ & $\lambda_2=4$ &$\lambda_3=3$\\ 
  
 && $\chi_1=0$ &$\chi_2=0$&$\chi_3=0$\\
  
 && $\psi_1=8$ &$\psi_2=8$&$\psi_3=6$\\
 && $p_1=0.4$ &$p_2=0.4$&$p_3=0.2$ \\
\hline
 Nonsense & \multicolumn{4}{c}{\multirow{2}{*}{}$\vecX_3\sim \text{GIG}(3,0,6)$}\\

 &\multicolumn{4}{c}{$\vecX_4 \sim \text{GIG}(1,0,2)$}\\ 
 \hline
 Noisy& \multicolumn{4}{c}{$\vecX_5 = 0.6\mathbf{V1}+0.4\mathbf{Z}, \ \text{where} \ \mathbf{Z} \sim N(0,5)$}\\
 \hline
\end{tabular}
\end{table}
\end{singlespace}
\begin{figure}[H]
    \centering
    \includegraphics[scale=0.95]{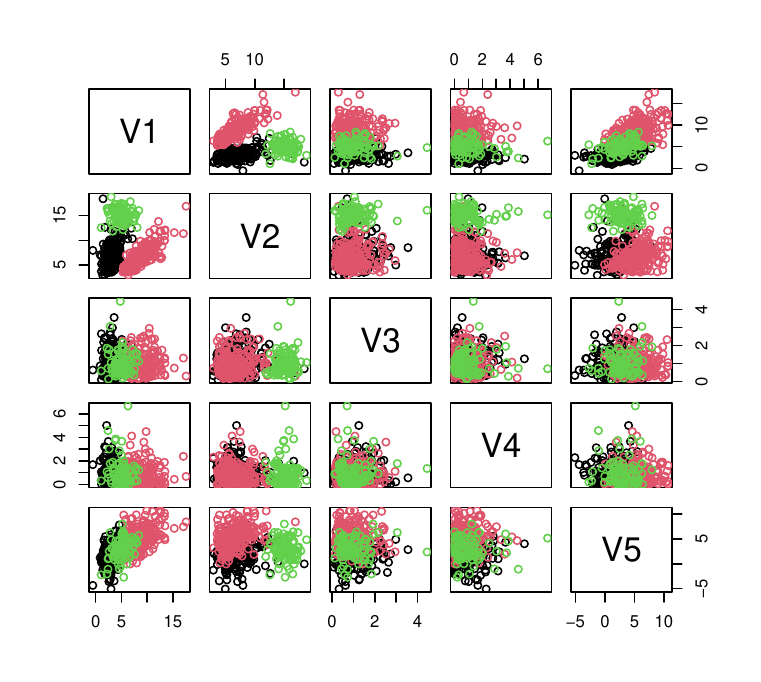}
    \vspace{-0.2in}
    \captionof{figure}{Example of simulated data when $N=500$.}
    \label{fig:simulatedD}
    \end{figure}
\begin{table}[H]
\centering
\caption{Summary of simulation results for all approaches. }
\label{tab:simres}
\begin{tabular}{ lcccccccccc } 
\hline
Method & $N$ & $\overline{G}$ & $\overline{\text{ARI}}$ (s.d.) & V1 & V2 & V3 & V4 &V5 \\
\hline
\multirow{3}{*}{} \vscc & 200 &  4.6 & 0.73 (0.14) &250&250&151&193&174\\ 
& 500 &  6.4 & 0.57 (0.1) & 250 &250 &232&249&192 \\ 
& 1000 & 7.8  & 0.44 (0.06) & 250&250&241&250&175\\ 
\hline
\multirow{3}{*}{} \clustvarsel & 200  & 4.6 & 0.68 (0.2) &243&244&10&138&0\\ 
& 500 & 7.4 & 0.46 (0.09) & 250 &250 &2&247&0 \\ 
& 1000 &  8.3 & 0.41 (0.05) & 250&250&18&250&0\\ 
\hline
 \rowcolor{Gray}
\multirow{3}{*}{} \vsccm-forwards & 200 & 3.02 & 0.85 (0.2) &234&240&32&18&39 \\ 
 \rowcolor{Gray}
& 500 &  3.7 & 0.83 (0.14) & 250&243&19&41&128\\
 \rowcolor{Gray}
& 1000 & 5 & 0.68 (0.17) & 248&227&8&78&104\\ 
\hline
 \rowcolor{Gray}
\multirow{3}{*}{} \vsccm-backwards & 200 & 2.9& 0.90 (0.14) & 239&249&20&18&19 \\ 
 \rowcolor{Gray}
& 500 & 3 & 0.94 (0.03) & 250&250&0&1&107\\ 
 \rowcolor{Gray}
& 1000 & 3 & 0.95 (0.01) & 250&250&0&0&140\\ 
\hline
 \rowcolor{Gray}
\multirow{3}{*}{} \vsccm-full & 200 &  2.9 & 0.90 (0.13) & 239&249&25&14&17\\ 
 \rowcolor{Gray}
& 500 &  3 & 0.94 (0.03)& 250&250&0&1&120 \\ 
 \rowcolor{Gray}
& 1000 & 3 & 0.95 (0.02) & 250&250&0&0&145 \\ 
\hline
\multirow{3}{*}{}  \skewvarsel + MSN & 200  &  2.89 & 0.81 (0.31) &218&218&0&32&3\\ 
  & 500 &  3.09 & 0.90 (0.14) & 245&245&0&5&0\\ 
& 1000 & 3.5 & 0.87 (0.1) & 250&250&0&0&0\\ 
\hline
\end{tabular}
\end{table}

\section{Discussion} \label{discuss}
In nearly all instances, we see the skewed extensions of common variable selection algorithms improving performance in the presence of skewness. This improvement in performance is seen not only in the selection of the number of clusters and ARI but, more importantly, in the reduction of dimensions. In the AIS and breast cancer examples, we see more effective dimension reduction by \vsccm\ than \skewvarsel\ in terms of the magnitude of dimension reduction and model fitting performance. For the banknote dataset, \vsccm\ selects more variables than \skewvarsel\ but also results in a better fitting model, regardless of the model fit to the \skewvarsel\ results. The Italian wine dataset highlights the potential importance of utilizing methods designed for Gaussian clusters when appropriate.

There are instances where \vsccm\ may select too many variables to account for some odd observations. For example, we see this in the AIS dataset when the backwards and full Manly mixture extensions select variable Hg. This selection causes some boundary points between groups to switch clusters resulting in one less miss-classification. Although adding this variable improves ARI, the goal of these algorithms is dimension reduction; as such, there may be instances in which smaller ARI is preferred if it corresponds to smaller number of selected variables. One way to account for odd or hard-to-classify observations may be mixtures of contaminated transformation distributions. These component densities contain an inflated secondary component that allows for better modelling of outliers and heavy tails. 

One downside to the skewed extensions is computational overhead. The \clustvarsel\ and \skewvarsel\ algorithms are naturally more computationally expensive due to their step-wise nature, with \skewvarsel\ taking longer as more parameters need to be estimated. The \vscc\ algorithm outperforms all methods on computational time as model fitting takes place only on the initial full set and the final sets of variables. This improvement in computational time extends into the skewed space when a full Manly mixture is fit to the data. However, the algorithm slows down greatly under forward or backward transformation parameter selection due to the introduction of some inclusion/exclusion steps. This increase in computational time is heavily influenced by the structure of the clusters and the selection process used. For heavily skewed data, \vsccm\ with backwards selection is much faster than its forwards counterpart. If only a few non-zero transformation parameters are necessary, \vsccm\ with forwards selection would be much faster. Although more time-consuming than \vscc\ or \vsccm-full, \vsccm\ with transformation parameter selection does tend to perform better than both in terms of ARI and dimension reduction. Thus, we suggest one performs some exploratory analysis on their data before selecting any of these methods to ensure that the algorithm selected is a good fit for their data and the computational overhead is justified. Additionally, the Manly transformation parameter selection is currently programmed in {\sf R} and could be sped up if programmed in a faster language. Computational time could be further reduced with parallelization of model fitting within each inclusion/exclusion step. 


\begin{thebibliography}{}
\vspace{-0.1in}
\bibitem[\protect\citeauthoryear{Adams and Beling}{Adams and
  Beling}{2019}]{adams2019}
Adams, S. and P.~A. Beling (2019).
\newblock A survey of feature selection methods for gaussian mixture models and
  hidden markov models.
\newblock {\em Artificial Intelligence Review\/}~{\em 52\/}(3), 1739--1779.
\vspace{-0.1in}
\bibitem[\protect\citeauthoryear{Andrews and McNicholas}{Andrews and
  McNicholas}{2014}]{andrews14}
Andrews, J.~L. and P.~D. McNicholas (2014).
\newblock Variable selection for clustering and classification.
\newblock {\em Journal of Classification\/}~{\em 31\/}(2), 136--153.
\vspace{-0.1in}
\bibitem[\protect\citeauthoryear{Loperfido}{Loperfido}{2019}]{loperfido19}
Loperfido, N. (2019).
\newblock Finite mixutres, projection pursuit and tensor rank: a triangulation.
\newblock {\em Advances in Data Analysis and Classification\/}~{\em 13\/}(1), 145--173.
\vspace{-0.1in}
\bibitem[\protect\citeauthoryear{Redner and Walker}{Redner and Walker}{1984}]{redner84}
Redner, R.~A. and H.~F. Walker (1984).
\newblock Mixture densities, maximum likelihood and the EM algorithm.
\newblock {\em SIAM Review\/}~{\em 26\/}, 195--239.
\vspace{-0.1in}
\bibitem[\protect\citeauthoryear{Stephens}{Stephens}{2000}]{stephens2000}
Stephens, M. (2000).
\newblock Dealing with label switching in mixture models.
\newblock {\em Journal of Royal Statistical Society: Series B\/}~{\em 62\/}, 795--809.
\vspace{-0.1in}
\bibitem[\protect\citeauthoryear{Andrews, Neal, and McNicholas}{Andrews
  et~al.}{2023}]{andrews22}
Andrews, J.~L., M.~Neal, and P.~D. McNicholas (2023).
\newblock {\em {vscc}: Variable Selection for Clustering and Classification}.
\newblock R package version 0.5.
\vspace{-0.1in}
\bibitem[\protect\citeauthoryear{Barndorff-Nielsen, Kent, and
  S{\o}rensen}{Barndorff-Nielsen et~al.}{1982}]{barndorff82}
Barndorff-Nielsen, O., J.~Kent, and M.~S{\o}rensen (1982).
\newblock Normal variance-mean mixtures and z distributions.
\newblock {\em International Statistical Review\/}~{\em 50\/}(2), 145--159.
\vspace{-0.1in}
\bibitem[\protect\citeauthoryear{Bouveyron and Brunet-Saumard}{Bouveyron and
  Brunet-Saumard}{2014}]{bouveyron14}
Bouveyron, C. and C.~Brunet-Saumard (2014).
\newblock Model-based clustering of high-dimensional data: A review.
\newblock {\em Computational Statistics and Data Analysis\/}~{\em 71}, 52--78.
\vspace{-0.1in}
\bibitem[\protect\citeauthoryear{Box and Cox}{Box and Cox}{1964}]{box1964}
Box, G.~E. and D.~R. Cox (1964).
\newblock An analysis of transformations.
\newblock {\em Journal of the Royal Statistical Society: Series B
  (Methodological)\/}~{\em 26\/}(2), 211--243.
\vspace{-0.1in}
\bibitem[\protect\citeauthoryear{Browne and McNicholas}{Browne and
  McNicholas}{2015}]{browne15}
Browne, R.~P. and P.~D. McNicholas (2015).
\newblock A mixture of generalized hyperbolic distributions.
\newblock {\em Canadian Journal of Statistics\/}~{\em 43\/}(2), 176--198.
\vspace{-0.1in}
\bibitem[\protect\citeauthoryear{Dua and Graff}{Dua and Graff}{2019}]{Dua2019}
Dua, D. and C.~Graff (2019).
\newblock {UCI} machine learning repository.
\vspace{-0.1in}
\bibitem[\protect\citeauthoryear{Fop and Murphy}{Fop and Murphy}{2018}]{fop18}
Fop, M. and T.~B. Murphy (2018).
\newblock Variable selection methods for model-based clustering.
\newblock {\em Statistics Surveys\/}~{\em 12}, 18--65.
\vspace{-0.1in}
\bibitem[\protect\citeauthoryear{Franczak, Browne, and McNicholas}{Franczak
  et~al.}{2014}]{franczak14}
Franczak, B.~C., R.~P. Browne, and P.~D. McNicholas (2014).
\newblock Mixtures of shifted asymmetric {L}aplace distributions.
\newblock {\em IEEE Transactions on Pattern Analysis and Machine
  Intelligence\/}~{\em 36\/}(6), 1149--1157.
\vspace{-0.1in}
\bibitem[\protect\citeauthoryear{Gallaugher, McNicholas, Melnykov, and
  Zhu}{Gallaugher et~al.}{2020}]{gallaugher20a}
Gallaugher, M. P.~B., P.~D. McNicholas, V.~Melnykov, and X.~Zhu (2020).
\newblock Skewed distributions or transformations? modelling skewness for a
  cluster analysis.
\vspace{-0.1in}
\bibitem[\protect\citeauthoryear{Hubert and Arabie}{Hubert and
  Arabie}{1985}]{hubert85}
Hubert, L. and P.~Arabie (1985).
\newblock Comparing partitions.
\newblock {\em Journal of Classification\/}~{\em 2\/}(1), 193--218.
\vspace{-0.1in}
\bibitem[\protect\citeauthoryear{Nelder and Mead}{Nelder and
  Mead}{1965}]{nelder65}
Nelder, J.A.. and R.~Mead (1965).
\newblock A Simplex Method for Function Minimization
\newblock {\em The Computer Journal\/}~{\em 7\/}(1), 308-313.
\vspace{-0.1in}
\bibitem[\protect\citeauthoryear{Hennig}{Hennig}{2023}]{hennig2023parameters}
Hennig, C. (2023).
\newblock Parameters not empirically identifiable or distinguishable, including correlation between Gaussian observations.
\newblock {\em Statistical Papers\/}~{\em 2\/}(1), 1--24.
\vspace{-0.1in}
\bibitem[\protect\citeauthoryear{Koot, Mandjes, van't Noordende, and De Laat}{Koot et~al.}{2013}]{koot2013probabilistic}
Koot, M., M.~Mandjes, G.~van't Noordende, and C.~De Laat (2013).
\newblock A probabilistic perspective on re-identifiability.
\newblock {\em Mathematical Population Studies\/}~{\em 20\/}(3), 155--171.
\vspace{-0.1in}
\bibitem[\protect\citeauthoryear{Karlis and Santourian}{Karlis and
  Santourian}{2009}]{karlis09}
Karlis, D. and A.~Santourian (2009).
\newblock Model-based clustering with non-elliptically contoured distributions.
\newblock {\em Statistics and Computing\/}~{\em 19\/}(1), 73--83.
\vspace{-0.1in}
\bibitem[\protect\citeauthoryear{Kass and Raftery}{Kass and
  Raftery}{1995}]{kass95}
Kass, R.~E. and A.~E. Raftery (1995).
\newblock Bayes factors.
\newblock {\em Journal of the American Statistical Association\/}~{\em
  90\/}(430), 773--795.
\vspace{-0.1in}
\bibitem[\protect\citeauthoryear{Lo and Gottardo}{Lo and
  Gottardo}{2012}]{lo2012}
Lo, K. and R.~Gottardo (2012).
\newblock Flexible mixture modeling via the multivariate t distribution with
  the box-cox transformation: an alternative to the skew-t distribution.
\newblock {\em Statistics and computing\/}~{\em 22\/}(1), 33--52.
\vspace{-0.1in}
\bibitem[\protect\citeauthoryear{Maugis, Celeux, and Martin-Magniette}{Maugis
  et~al.}{2009}]{maugis09a}
Maugis, C., G.~Celeux, and M.-L. Martin-Magniette (2009).
\newblock Variable selection for clustering with {G}aussian mixture models.
\newblock {\em Biometrics\/}~{\em 65\/}(3), 701--709.
\vspace{-0.1in}
\bibitem[\protect\citeauthoryear{McNicholas}{McNicholas}{2016a}]{mcnicholas16a}
McNicholas, P.~D. (2016a).
\newblock {\em Mixture Model-Based Classification}.
\newblock Boca Raton: Chapman \& Hall/CRC Press.
\vspace{-0.1in}
\bibitem[\protect\citeauthoryear{McNicholas}{McNicholas}{2016b}]{mcnicholas16b}
McNicholas, P.~D. (2016b).
\newblock Model-based clustering.
\newblock {\em Journal of Classification\/}~{\em 33\/}(3), 331--373.
\vspace{-0.1in}
\bibitem[\protect\citeauthoryear{McNicholas, ElSherbiny, McDaid, and
  Murphy}{McNicholas et~al.}{2022}]{mcnicholas2022}
McNicholas, P.~D., A.~ElSherbiny, A.~F. McDaid, and T.~B. Murphy (2022).
\newblock {\em pgmm: Parsimonious Gaussian Mixture Models}.
\newblock R package version 1.2.6.
\vspace{-0.1in}
\bibitem[\protect\citeauthoryear{McNicholas, McNicholas, and Browne}{McNicholas
  et~al.}{2017}]{smcnicholas17}
McNicholas, S.~M., P.~D. McNicholas, and R.~P. Browne (2017).
\newblock A mixture of variance-gamma factor analyzers.
\newblock In S.~E. Ahmed (Ed.), {\em Big and Complex Data Analysis:
  Methodologies and Applications}, pp.\  369--385. Cham: Springer International
  Publishing.
\vspace{-0.1in}
\bibitem[\protect\citeauthoryear{Pyne, Hu, Wang, Rossin, Lin, Maier,
  Baecher-Allan, McLachlan, Tamayo, Hafler, Jager, and Mesirow}{Pyne
  et~al.}{2009}]{pyne09}
Pyne, S., X.~Hu, K.~Wang, E.~Rossin, T.-I. Lin, L.~M. Maier, C.~Baecher-Allan,
  G.~J. McLachlan, P.~Tamayo, D.~A. Hafler, P.~L.~D. Jager, and J.~Mesirow
  (2009).
\newblock Automated high-dimensional flow cytometric data analysis.
\newblock {\em Proceedings of the National Academy of Sciences\/}~{\em 106},
  8519--8524.
\vspace{-0.1in}
\bibitem[\protect\citeauthoryear{{R Core Team}}{{R Core Team}}{2023}]{R23}
{R Core Team} (2023).
\newblock {\em R: A Language and Environment for Statistical Computing}.
\newblock Vienna, Austria: R Foundation for Statistical Computing.
\vspace{-0.1in}
\bibitem[\protect\citeauthoryear{Raftery and Dean}{Raftery and
  Dean}{2006}]{raftery06}
Raftery, A.~E. and N.~Dean (2006).
\newblock Variable selection for model-based clustering.
\newblock {\em Journal of the American Statistical Association\/}~{\em
  101\/}(473), 168--178.
\vspace{-0.1in}
\bibitem[\protect\citeauthoryear{Rand}{Rand}{1971}]{rand71}
Rand, W.~M. (1971).
\newblock Objective criteria for the evaluation of clustering methods.
\newblock {\em Journal of the American Statistical Association\/}~{\em
  66\/}(336), 846--850.
\vspace{-0.1in}
\bibitem[\protect\citeauthoryear{Schwarz}{Schwarz}{1978}]{schwarz78}
Schwarz, G. (1978).
\newblock Estimating the dimension of a model.
\newblock {\em The Annals of Statistics\/}~{\em 6\/}(2), 461--464.
\vspace{-0.1in}
\bibitem[\protect\citeauthoryear{Scrucca, Fop, Murphy, and Raftery}{Scrucca
  et~al.}{2016}]{scrucca16}
Scrucca, L., M.~Fop, T.~B. Murphy, and A.~E. Raftery (2016).
\newblock {mclust} 5: clustering, classification and density estimation using
  {G}aussian finite mixture models.
\newblock {\em The {R} Journal\/}~{\em 8\/}(1), 289--317.
\vspace{-0.1in}
\bibitem[\protect\citeauthoryear{Scrucca and Raftery}{Scrucca and
  Raftery}{2018}]{scrucca18}
Scrucca, L. and A.~E. Raftery (2018).
\newblock clustvarsel: A package implementing variable selection for gaussian
  model-based clustering in r.
\newblock {\em Journal of Statistical Software\/}~{\em 84\/}(1), 1–28.
\vspace{-0.1in}
\bibitem[\protect\citeauthoryear{Steinley and Brusco}{Steinley and
  Brusco}{2008}]{steinley08}
Steinley, D. and M.~J. Brusco (2008).
\newblock Selection of variables in cluster analysis: An empirical comparison
  of eight procedures.
\newblock {\em Psychometrika\/}~{\em 73}, 125--144.
\vspace{-0.1in}
\bibitem[\protect\citeauthoryear{Steinley and Brusco}{Steinley and
  Brusco}{2011}]{steinley11b}
Steinley, D. and M.~J. Brusco (2011).
\newblock K-means clustering and mixture model clustering: Reply to mclachlan
  (2011) and vermunt (2011).
\newblock {\em Psychological Methods\/}~{\em 16\/}(1), 89--92.
\vspace{-0.1in}
\bibitem[\protect\citeauthoryear{Wallace, Buysse, Germain, Hall, and
  Iyengar}{Wallace et~al.}{2018}]{wallace18}
Wallace, M.~L., D.~J. Buysse, A.~Germain, M.~H. Hall, and S.~Iyengar (2018).
\newblock Variable selection for skewed model-based clustering: application to
  the identification of novel sleep phenotypes.
\newblock {\em Journal of the American Statistical Association\/}~{\em
  113\/}(521), 95--110.
\vspace{-0.1in}
\bibitem[\protect\citeauthoryear{Zhu and Melnykov}{Zhu and
  Melnykov}{2018a}]{zhu18}
Zhu, X. and V.~Melnykov (2018a).
\newblock Manly transformation in finite mixture modeling.
\newblock {\em Computational Statistics \& Data Analysis\/}~{\em 121},
  190--208.
\vspace{-0.1in}
\bibitem[\protect\citeauthoryear{Zhu and Melnykov}{Zhu and
  Melnykov}{2018b}]{zhu18a}
Zhu, X. and V.~Melnykov (2018b).
\newblock {\em {ManlyMix}: An {R} Package for Model-Based Clustering with
  {M}anly Mixture Models}.
\newblock R package version 0.1.14.

\end{thebibliography}

\subsection*{Acknowledgements}
This work is supported by a Discovery Grant from the Natural Sciences and Engineering Research Council of Canada, a Dorothy Killam Fellowship, and the Canada Research Chairs program.

\end{document}